# Thermorheological Complexity in Polymers and the Problem of the Glass Transition

K.L. Ngai[1] and C.M. Roland[2]

[1] *Dipartimento di Fisica, Università di Pisa, Largo B. Pontecorvo 3,I-56127, Pisa, Italy*

[2] *Naval Research Laboratory, Washington DC 20375-5342*

ABSTRACT: A current focus in studies of the glass transition is the role of dynamic heterogeneities. Although these efforts may clarify the origin of the spectacular change in properties of liquids approaching vitrification, we point out that a seemingly related phenomenon – thermorheological complexity in polymers – must involve different mechanisms. In particular, as seen from consideration of various properties involving the chain dynamics, averaging over different length and time scales cannot offer a resolution to the problem of thermorheological complexity.

---

The "glass transition problem", which refers to the spectacular change in dynamic properties of liquids cooled towards their glass transition temperature, remains unsolved after more than a half-century of research. The difficulty is that there are many properties associated with vitrification of a liquid, and these properties have interrelationships which complicate any analysis. Recently attention has been directed at the spatial heterogeneity of the dynamics; this "dynamic heterogeneity" is a general feature of supercooled liquids that has been investigated both experimentally and by computer simulations. Whether dynamic heterogeneity will turn out to be the key to resolving the glass transition problem remains to be seen. Our interest herein is a seemingly related problem – the thermorheological complexity of polymers. This refers to the different temperature dependences of the various modes of motion in polymers, which leads to a breakdown of the time-temperature superposition principle. In fact, the differing response of global and local motions in polymers extends to various thermodynamic quantities (Figure 1 [1]), the failure of time-temperature superpositioning being just the most often observed manifestation of the phenomenon.

There is a common belief that dynamic heterogeneity underlies this thermorheological complexity, and that upon resolution of the glass transition problem, further investigation of the polymer problem will no longer be required. This idea arises from the notion that the different temperature-dependences of the segmental and chain motions is due to dynamic heterogeneities affecting the former but being averaged out over the longer length and time scales of the latter [2,3]. The purpose of this short note is to point out that neither dynamic heterogeneity as currently applied nor an eventual solution of the glass transition problem can likely provide an explanation for the thermorheological complexity of polymers.

The traditional assumption in the field of polymers is that all viscoelastic mechanisms over all length-scales are governed by the same monomeric friction coefficient, and thus have the same dependence on temperature, pressure, etc. This ideal behavior leads to time-temperature superpositioning, and is inherent to classic theories of polymer dynamics such as the Rouse and reptation models [4,5]. Deviation from the single friction factor idea was first observed in creep compliance measurements for entangled polymers [6], and subsequently, for the segmental relaxation and longer length-scale chain relaxation in unentangled polymers [7]. These findings have since been confirmed in a large number of materials [4,8,9], so that thermorheological complexity can be considered a general property of polymers. The fact that this behavior cannot be directly connected to dynamic heterogeneity and the glass transition problem is indicated by several experimental facts:

1. The self-diffusion constant and viscosity of entangled polymers are decoupled, with the apparent activation energy for self-diffusion being significantly smaller than for the viscosity [10]. This difference is similar to the difference in temperature dependences for the segmental and chain dynamics. Since the self-diffusion constant and the viscosity of entangled polymers *both involve global motions* (entailing length scales large compared to the dynamic heterogeneities), spatially heterogeneous dynamics cannot be the primary origin of the decoupling. That is, distinct averaging over the length and time scales cannot underlie differences between properties that involve similar length and time scales.
2. The viscosity of polymers has a different temperature dependence than that of the relaxation times governing the chain dynamics [8,11]. Depending on the molecular weight of the polymer and the measurement temperature, the viscosity can exhibit a stronger temperature dependence than even the local segmental modes. Such behavior cannot be explained as a consequence of averaging over different length scales, since the viscosity and chain relaxation are both global processes. Moreover, use of the averaging argument to explain the stronger temperature dependence of the local segmental dynamics obviates its use to explain a weaker temperature dependence.
3. The diminished role of dynamic heterogeneity for polymers is also evident in the substantial decrease of the steady state recoverable compliance, $J_s^0$, of low molecular weight polymers on cooling towards $T_g$ [8,12]. The product of this compliance and the viscosity yields the terminal relaxation time, so that the changes of $J_s^0$ with temperature cause another failure of the "single time scale" picture that cannot be ascribed to the spatial heterogeneity of the segmental dynamics.

4. The difference in temperature variations of chain and segmental friction has been observed mostly by mechanical compliance and modulus measurements, and occasionally by dielectric relaxation, but not from chain diffusion measurements. Thus, the argument [13] that dynamic heterogeneity is averaged out for chain diffusion in a different manner than for the segmental dynamics cannot obviously be applied to the problem of thermorheological complexity. This argument has been used to account for the breakdown of the Stokes-Einstein relation in molecular glass-formers, but this has been contradicted by both experiments and simulations [3].
5. The heterogeneous dynamics have been quantified for various materials, and for 1,4-polyisoprene the number of dynamically correlated repeat units, $N_c$, is on the order of 100, depending only weakly on the chain length [1]. If dynamic heterogeneity were related to thermorheological complexity, the expectation is that the latter would be magnified for polyisoprenes having a degree of polymerization significantly larger than $N_c$, while chains shorter than this value would be thermorheologically simple. However, the breakdown of time-temperature superpositioning has been reported for polyisoprenes having chain lengths that span $N_c$ [114,15,16,17]. There is no suggestion in these data of any connection between the length scale of the heterogeneous dynamics and the spatial dimensions of the chains.

As pointed out in a recent review [3], a complete understanding of supercooled liquids requires multiple interrelated lines of inquiry. For polymer scientists the task is even more daunting – neither dynamic heterogeneity as currently applied nor an eventual solution of the glass transition problem will explain the thermorheological complexity common to long chain molecules. Determining the mechanisms giving rise to different friction factors for different modes of motion in polymers is of fundamental importance, although it is neglected by mainstream viscoelastic theories. Even though a resolution to the glass transition problem through concepts based on dynamic heterogeneity may be tantalizingly near, thermorheological complexity will remain unsolved because it is a different issue involving different mechanisms.

**Acknowledgements**

The work at NRL was supported by the Office of Naval Research.

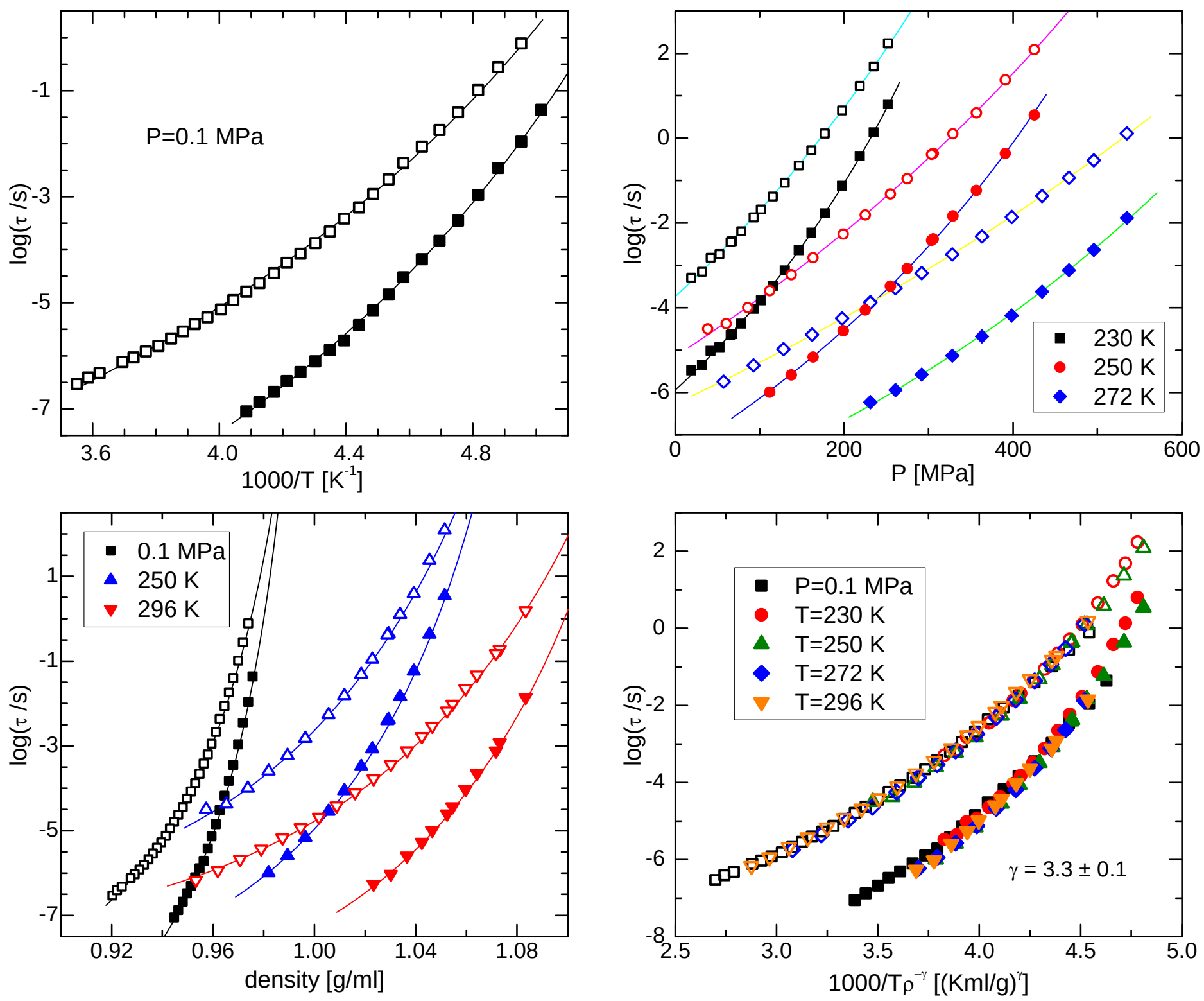

Figure 1. Relaxation times for the normal mode (open symbols) and local segmental dynamics (solid symbols) of unentangled 1,4-polyisoprene, as a function of inverse temperature, pressure, density, and the ratio of temperature to density with the latter raised to a material constant (respectively, from upper left to lower right). For all thermodynamic quantities, the segmental relaxation times exhibit a stronger dependence than does the normal mode. Data from ref. [1].